\documentstyle[amssymb,twoside]{article}
\input{tcilatex}

\begin{document}

\begin{center}
{\Large {\bf The Limits of Horn Logic Programs}} \footnote{{\small The
project is supported by the National 973 Project of China under the grant
number G1999032701 and the National Science Foundation of China.}}\\[0.5cm]
{Shilong Ma$^{\dag }$\footnote[2]{{\small Email: \{slma,
kexu\}@nlsde.buaa.edu.cn.}}, Yuefei Sui$^{\ddag }$\footnote[3]{{\small %
Email: suiyyff@sina.com.}}, and Ke Xu$^{\dag ,2}$}\\[0pt]
{\ {}$^{\dag }$ Department of Computer Science \\[0pt]
Beijing University of Aeronautics and Astronautics\\[0pt]
Beijing 100083, China}\\[0pt]
{\ {}$^{\ddag }$ Institute of Computing Technology, Academia Sinica \\[0pt]
Beijing 100080, China}
\end{center}

\bigskip

\begin{minipage}{11cm} 
{\small\footnotesize {\bf   Abstract}: \ Given a sequence $\{\Pi_n\}$ of Horn logic 
programs, the limit $\Pi$ of $\{\Pi_n\}$ is the set of the clauses such that 
every clause in $\Pi$ belongs to almost every $\Pi_n$ and every clause in 
infinitely many $\Pi_n$'s belongs to $\Pi$ also. The limit program $\Pi$ is still Horn but may be infinite.
In this paper, we consider if 
the least Herbrand model of the limit of a given Horn logic program 
sequence $\{\Pi_n\}$ equals the limit of the least Herbrand models of each logic program $\Pi_n$. 
It is proved that this property is not true in general but holds if under an assumption which can 
be syntactically checked and be satisfied by a  
class of Horn logic programs. Thus, under this assumption we can approach the least Herbrand model of 
the limit $\Pi$ by the sequence of the least Herbrand models of each finite program $\Pi_n$. 
We also prove that if a finite Horn logic program 
satisfies this assumption, then the least Herbrand model of this program is 
recursive. Finally, by use of the concept of stability from dynamical systems, 
we prove that this assumption is exactly a sufficient condition to guarantee 
the stability of fixed points for Horn logic programs. 

\linebreak Decidability.}
\end{minipage}

\bigskip

\noindent{\bf 1.Introduction}

\smallskip

As time goes by, the knowledge of the human being is increasing
exponentially. The amount of information on the Internet is doubled in
several months. So it becomes more and more important to discover useful
knowledge in massive information. There have been several data mining tools
to find useful knowledge from very large databases. If the knowledge found
by a data mining tool is taken as a theory at the current time, we assume
that the theory is consistent. As the knowledge increases, the theories
should also be updated. Thus we get a sequence of theories $\Pi _{1},\Pi
_{2},\cdots ,\Pi _{n},\cdots $. This procedure may never stop, i.e. maybe
there does not exist a natural number $k$ such that $\Pi _{k}=\Pi
_{k+1}=\cdots $. For example, if we restrict the theories to Horn logic
programs, then there exists some Herbrand interpretation $I$ such that we
will never find a finite program $\Pi $ whose least Herbrand model $M=I,$
because that the set of Herbrand interpretations is uncountable while the
set of finite programs is only countable$^{[8]}$. So sometimes we need to
consider some kind of limit of theories and discover what kind of knowledge
is true in the limit.

Formally introducing limits of sequences of first order theories into logic
and computer science, and using theory versions as approximations of some
formal theories in convergent infinite computations are independent
contributions by Li in 1992. Li$^{[5],[6],[7]}$ first defined the limits of
first order theories, and thereon gave a formal system of the inductive
logic. Precisely, given a sequence $\{\Pi _{n}\}$ of the first order
theories $\Pi _{n}$'s, the limit $\Pi =\displaystyle\lim_{n\rightarrow
\infty }\Pi _{n}$ is the set of the sentences such that every sentence in $%
\Pi $ belongs to almost every $\Pi _{n},$ and every sentence in infinitely
many $\Pi _{n}$'s belongs to $\Pi $ also. The limit does not always exist
for any sequence of the first order theories.

We shall consider the limits of the logic programs in terms of their
Herbrand models, and focus on the Horn logic programs. There are two reasons
for us to focus on the Horn logic programs. Firstly, the Horn logic programs
are used as a main representation method of knowledge in the knowledge
engineering. Every knowledge base must have the ability to reason, and the
reasoning in a Horn logic program is feasible. Secondly, the Horn logic
programs have the least Herbrand model$^{[10]}$, since the meet of any two
Herbrand models of a Horn logic program is a Herbrand model of this Horn
logic program.

For a sequence $\{\Pi _{n}\}$ of finite Horn logic programs, if the limit $%
\Pi $ of $\{\Pi _{n}\}$ exists then $\Pi $ is a Horn logic program but it
may be infinite. To discover what is true in $\Pi $,\ it is crucial to
compute the least Herbrand model of $\Pi $. Then, the problem is: {\sl How to
construct the least Herbrand models of such Horn logic programs}?

We know that for every finite $\Pi _{n}$, the least Herbrand model can be
constructed. Therefore, one may naturally wonder if the least Herbrand model
of $\Pi $ can be approached by the sequence of the least Herbrand models of $%
\Pi _{n}$. Let $M_{n}$ and $M$ be the least Herbrand models of $\Pi _{n}$
and $\Pi $ respectively. Namely, we hope to have 
$$
M=\displaystyle\lim_{n\rightarrow \infty }M_{n}.\eqno(\ast ) 
$$%
In another form, ${\bf M}(\displaystyle\lim_{n\rightarrow \infty }\Pi _{n})=%
\displaystyle\lim_{n\rightarrow \infty }{\bf M}(\Pi _{n}),$ where ${\bf M}$
is taken as an operator which maps a Horn logic program to the least
Herbrand model of the logic program.

In this paper, we prove that $(\ast )$ holds for a class of Horn logic
programs. Thus, if  $\{\Pi _{n}\}$ is a sequence of finite Horn logic
programs in this class and $\Pi =\displaystyle\lim_{n\rightarrow \infty }\Pi
_{n}$ exists, then we can approach the least Herbrand model of $\Pi $ by the
sequence of the least Herbrand models of each program $\Pi _{n}$.

The paper is arranged as follows. In section 2, we will give the basic
definitions in logic programs, and the fixpoint semantics of Horn logic
programs. In section 3, we will give two examples to show that $(\ast )$ is
not true in general but holds for some Horn logic programs. In section 4, we
will prove that $(\ast )$ is true under an assumption which can be syntactically 
checked and easily satisfied, and if a finite Horn
logic program satisfies this assumption then the least Herbrand model of
this program is recursive. In section 5, by use of the concept of stability
from dynamical systems, we will prove that the assumption in section 4 is
exactly a sufficient condition to guarantee the stability of fixed points
for Horn logic programs.

Our notation follows from Dahr's book$^{[2]}.$ We use $\Pi $ to denote a
logic program, $p,q$ atoms, $\pi $ a clause and $\Theta $ a substitution. $%
I,J$ denote subsets of Herbrand bases, or interpretations; $n,m,r,s,N$
denote the natural numbers. \bigskip

\noindent {\bf 2. The basic definitions}

\smallskip

Let the base (language) $L$ consist of finite sets of constant symbols,
variable symbols, function symbols and predicate symbols. A term or formula
is {\it ground}\ if there is no variable in it. Let $TE_{L}$ be the set of
all terms in $L$, and $TG_{L}$ of all the ground terms in $L.$ A {\it literal%
} is an atom or the negation of an atom. An atom is called a {\it positive}
literal whereas a negated atom is called a {\it negative} literal. A clause $%
\pi :p_{1}\vee \cdots \vee p_{r}\leftarrow q_{1},\cdots ,q_{s}$ is called a 
{\it rule}\thinspace\ if $r\geq 1$ and $s\geq 1;$ a {\it fact}\thinspace\ if 
$r\geq 1$ and $s=0;$ and a {\it goal}\thinspace\ if $r=0$ and $s\geq 1.\
p_{1}\vee \cdots \vee p_{r}$ is called the {\it rule head}, denoted by $%
head(\pi ),$ and $q_{1},\cdots ,q_{s}$ is called the {\it rule body},
denoted by $body(\pi ).$ Every $q_{i}$ is called a {\it subgoal.} We assume
that the clauses are closed, i.e., every clause is a closed formula with
universal quantification on variables in the formula, for example, $%
p(x)\leftarrow q(x)$ means that $\forall x[p(x)\leftarrow q(x)].$ A {\it %
logic program}\ is a set of clauses in some base $L.$

{\bf Definition 2.1.} A clause $\pi :p_{1}\vee \cdots \vee p_{r}\leftarrow
q_{1},\cdots ,q_{s}$ is called {\it Horn clause} if $r\leq 1$ and every
subgoal is an atom. A logic program $\Pi $ is {\it Horn}\ if every formula
in $\Pi $ is a Horn clause.

The {\it Herbrand universe} $U_L$ of a logic program in $L$ is the set of
all ground terms in $L.\ HB_L$ denotes the set of all the ground atoms in $%
L, $ called the {\it Herbrand base}.

{\bf Definition 2.2.} A {\it Herbrand interpretation}\ of a logic program $%
\Pi$ is any subset $I$ of $HB.$ A Herbrand interpretation $I$ of $\Pi$ is a 
{\it Herbrand model of} $\Pi$ if every clause is satisfied under $I.$ 
%Let $\Pi$ be a set of clauses; $S$ has a model if and only if $S$ has a Herbrand model.
A Herbrand model $I$ of a logic program $\Pi$ is called {\it minimal} if
there exists no subset $I^{\prime}\subseteq I$ that is a model for $\Pi;$ 
{\it least model} if $I$ is the unique minimal model of $\Pi.$

{\bf Theorem 2.3}. ([10]) Let $\Pi $ be a Horn program and let $HM(\Pi )$
denote the set of all Herbrand models of $\Pi .$ Then the model intersection
property holds, i.e., $\displaystyle\bigcap_{W\in HM(\Pi )}W$ is a Herbrand
model of $\Pi .$

\bigskip

Let $S$ be a nonempty set and $\wp (S)$ be the power set of $S.$ Then $(\wp
(S),\subseteq )$ be a complete lattice. A mapping $f:\wp (S)\rightarrow \wp
(S)$ is {\it monotonic}\ if for all element $a,b\in \wp (S),\ a\subseteq b$
implies $f(a)\subseteq f(b);$ $f$ is {\it finitary}\ if $f(\bigcup_{n=0}^{%
\infty }a_{n})\subseteq \bigcup_{n=0}^{\infty }f(a_{n})$ for every infinite
sequence $a_{0}\subseteq a_{1}\subseteq \cdots ;\ f$ is {\it continuous}\ if
it is monotonic and finitary.

{\bf Definition 2.4.} A subset $a\subseteq \wp(S)$ is called a {\it %
pre-fixpoint}\ if $f(a)\subseteq a;$ a {\it post-fixpoint}\ if $a\subseteq
f(a);$ and a {\it fixpoint}\ if $a=f(a).$

Let $f:\wp(S)\rightarrow \wp(S)$ be a monotonic mapping. Then $f$ has a
least fixpoint ${\rm lfp(f).}$ Define by induction on $n$ the following
elements of $\wp(S):$ 
\[
\begin{array}{rl}
f^0 & =\emptyset; \\ 
f^1 & =f(\emptyset)=f(f^0); \\ 
f^{n+1} & =f(f^n).%
\end{array}%
\]
Let $f^\omega=\bigcup_{n=0}^\infty f^n.$

{\bf Theorem 2.5}(Kleene).\ Let $f:\wp (S)\rightarrow \wp (S)$ be a
continuous mapping. Then ${\rm lfp(f)=f^{\omega }.}$

\bigskip

In 1992, Li$^{[5]}$ defined the limit of a theory sequence using the
standard set-theoretic definition of limit. By this definition, given a
sequence $\{\Pi _{n}\}$ of logic programs, we say that 
\[
\overline{\displaystyle\lim }_{n\rightarrow \infty }\Pi _{n}=\displaystyle%
\bigcap_{i=1}^{\infty }\displaystyle\bigcup_{j=i}^{\infty }\Pi _{j} 
\]%
is the upper limit of $\{\Pi _{n}\},$ and 
\[
\underline{{\displaystyle\lim }}_{n\rightarrow \infty }\Pi _{n}=\displaystyle%
\bigcup_{i=1}^{\infty }\displaystyle\bigcap_{j=i}^{\infty }\Pi _{j} 
\]%
is the lower limit of $\{\Pi _{n}\}.$ It follows from the definition of
limits that given a sequence $\{\Pi _{n}\}$ of logic programs, a clause
belongs to the upper limit iff it belongs to infinitely many $\Pi _{n}$'s,
and a clause belongs to the lower limit iff it belongs to almost every $\Pi
_{n}$. It is not hard to prove that $\underline{\displaystyle\lim }%
_{n\rightarrow \infty }\Pi _{n}\sqsubseteq \overline{\displaystyle\lim }%
_{n\rightarrow \infty }\Pi _{n}$. If $\overline{\displaystyle\lim }%
_{n\rightarrow \infty }\Pi _{n}=\underline{\displaystyle\lim }_{n\rightarrow
\infty }\Pi _{n}$ then we say that the (set-theoretic) limit of $\{\Pi
_{n}\} $ exists and denote it by $\displaystyle\lim_{n\rightarrow \infty
}\Pi _{n}.$

\bigskip

\noindent{\bf 3. The limits of the Horn logic programs}

\smallskip

Let $\Pi $ be a logic program. We define a mapping $f_{\Pi
}:2^{HB}\rightarrow 2^{HB}$ by, for any $I\in 2^{HB},$ 
\[
f_{\Pi }(I)=\{head(\pi )\Theta :\pi \in \Pi \ \&\ \exists \Theta \lbrack
body(\pi )\Theta \subseteq I]\}. 
\]

{\bf Proposition 3.1}([10]).\ If $\Pi $ is a Horn logic program then $f_{\Pi
}$ is monotonic and finitary. Moreover, $f_{\Pi }^{\omega }$ is the least
model of $\Pi .$

\bigskip

We hope to prove the following statement:

$(3.1)$\ \ Given a sequence $\{\Pi_n\}$ of the Horn logic programs, let $M_n$
be the least model of $\Pi_n.$ If $\Pi=\displaystyle\lim_{n\rightarrow
\infty} \Pi_n$ exists then $\Pi$ is a Horn logic program. Let $M$ be the
least model of $\Pi,$ then $M=\displaystyle\lim_{n\rightarrow \infty} M_n.$

The Claim $(3.1)$ does not hold for all Horn logic programs. Let us take a
look at the following example.

\bigskip

{\bf Example 3.2.}\ Assume that there are one function symbol $f,$ one
predicate symbol $p$ and one constant symbol $a$ in $L.$ Let $\{\Pi _{n}\}$
be defined as follows: 
\[
\begin{array}{rl}
\Pi _{1} & =\{p(x)\leftarrow p(f(x)),p(f(a))\}, \\ 
\Pi _{2} & =\{p(x)\leftarrow p(f(x)),p(f^{2}(a))\}, \\ 
& \cdots \\ 
\Pi _{n} & =\{p(x)\leftarrow p(f(x)),p(f^{n}(a))\},%
\end{array}%
. 
\]%
where $f^{(n)}(a)=f(f^{(n-1)}(a)),f^{1}(a)=f(a).$ Then we have 
\[
\begin{array}{rl}
f_{\Pi _{1}}^{\omega } & =\{p(a),p(f(a))\}, \\ 
f_{\Pi _{2}}^{\omega } & =\{p(a),p(f(a)),p(f^{2}(a))\}, \\ 
& \cdots \\ 
f_{\Pi _{n}}^{\omega } & =\{p(a),p(f(a)),\cdots ,p(f^{n}(a))\},%
\end{array}%
\]%
and 
\[
\displaystyle\lim_{n\rightarrow \infty }f_{\Pi _{n}}^{\omega
}=\{p(a),p(f(a)),\cdots ,p(f^{n}(a)),\cdots \}. 
\]%
But 
\[
\Pi =\displaystyle\lim_{n\rightarrow \infty }\Pi _{n}=\{p(x)\leftarrow
p(f(x))\}, 
\]%
and 
\[
f_{\Pi }^{\omega }=\emptyset . 
\]%
Therefore, $f_{\Pi }^{\omega }\neq \displaystyle\lim_{n\rightarrow \infty
}f_{\Pi _{n}}^{\omega }.$

\bigskip

For some other Horn logic programs, the Claim $(3.1)$ is true. Let us take a
look at the following example.

{\bf Example 3.3.}\ Assume that there are one function symbol $f,$ one
predicate symbol $p$ and one constant symbol $a$ in $L.$ Let $\{\Pi _{n}\}$
be defined as follows: 
\[
\begin{array}{rl}
\Pi _{1} & =\{p(f(x))\leftarrow p(x),p(f(a))\}, \\ 
\Pi _{2} & =\{p(f(x))\leftarrow p(x),p(f^{2}(a))\}, \\ 
& \cdots \\ 
\Pi _{n} & =\{p(f(x))\leftarrow p(x),p(f^{n}(a))\},%
\end{array}%
. 
\]%
where $f^{n}(a)=f(f^{n-1}(a)),f^{1}(a)=f(a).$ Then we have 
\[
\begin{array}{rl}
f_{\Pi _{1}}^{\omega } & =\{p(f(a)),p(f^{2}(a)),\cdots ,p(f^{m}(a)),\cdots
\}, \\ 
f_{\Pi _{2}}^{\omega } & =\{p(f^{2}(a)),p(f^{3}(a)),\cdots
,p(f^{m}(a)),\cdots \}, \\ 
& \cdots \\ 
f_{\Pi _{n}}^{\omega } & =\{p(f^{n}(a)),p(f^{n+1}(a)),\cdots \},%
\end{array}%
\]%
and 
\[
\displaystyle\lim_{n\rightarrow \infty }f_{\Pi _{n}}^{\omega }=\emptyset . 
\]%
Now we have 
\[
\Pi =\displaystyle\lim_{n\rightarrow \infty }\Pi _{n}=\{p(f(x))\leftarrow
p(x)\}, 
\]%
and 
\[
f_{\Pi }^{\omega }=\emptyset . 
\]%
Therefore, $f_{\Pi }^{\omega }=\displaystyle\lim_{n\rightarrow \infty
}f_{\Pi _{n}}^{\omega }.$

\bigskip \bigskip

\noindent{\bf 4. The main theorems}

\smallskip

Example 3.2 shows that the Claim $(3.1)$ does not hold for all Horn logic
programs. In general cases, we have the following property.

{\bf Lemma 4.1.} For any base $L,$ given any sequence $\{\Pi _{n}\}$ of the
Horn logic programs in $L,$ if $\Pi =\displaystyle\lim_{n\rightarrow \infty
}\Pi _{n}$ exists then $\underline{\displaystyle\lim }_{n\rightarrow \infty
}f_{\Pi _{n}}^{\omega }$ is a Herbrand model of $\Pi .$ Therefore, $f_{\Pi
}^{\omega }\subseteq \underline{\displaystyle\lim }_{n\rightarrow \infty
}f_{\Pi _{n}}^{\omega }.$

{\sl Proof.} To show that $f_{\Pi }^{\omega }\subseteq \underline{%
\displaystyle\lim }_{n\rightarrow \infty }f_{\Pi _{n}}^{\omega },$ let $p$
be a ground atom $\in f_{\Pi }^{\omega },$ we prove the claim by induction
on $m$ such that $p\in f_{\Pi }^{m+1}-f_{\Pi }^{m}.$ There are two cases:

Case 1. $p\in f_{\Pi }^{1}.$ Then $[p\leftarrow ]\in \Pi .$ Hence, for
almost every $n,[p\leftarrow ]\in \Pi _{n},$ so $p\in f_{\Pi
_{n}}^{1}\subseteq f_{\Pi _{n}}^{\omega }.$ Hence, $p\in \underline{%
\displaystyle\lim }_{n\rightarrow \infty }f_{\Pi _{n}}^{\omega }.$

Case 2. $p\in f_{\Pi }^{m+1}-f_{\Pi }^{m}.$ By the definition of $f_{\Pi },$
there is a clause $\pi \in \Pi $ and a substitution $\Theta $ such that $%
body(\pi )\Theta \subseteq f_{\Pi }^{m}$ and $p=head(\pi )\Theta .$ By the
induction assumption, $body(\pi )\Theta \in \underline{\displaystyle\lim }%
_{n\rightarrow \infty }f_{\Pi _{n}}^{\omega },$ i.e., there is an $N_{0}$
such that for any $n\geq N_{0},$ $body(\pi )\Theta \subseteq f_{\Pi
_{n}}^{\omega }.$ Since $\pi \in \Pi ,$ then $\pi \in \Pi _{n}$ for almost
every $n.$ Let $N_{1}$ be the least such that for any $n\geq N_{1},\ \pi \in
\Pi _{n}.$ Let $N=\max \{N_{0},N_{1}\}.$ Hence, by the same substitution $%
\Theta ,$ $p=head(\pi )\Theta \in f_{\Pi _{n}}^{\omega }$ for any $n\geq N.$
\hfill$\Box $

\bigskip

Example 3.2 and lemma 4.1 indicate that to make the Claim $(3.1)$ hold, we
have to put some conditions on Horn logic programs.

{\bf Assumption 4.2.} \ Given a Horn logic program $\Pi $, assume that every
clause $\pi :p\leftarrow p_{1},\cdots ,p_{m}$ in $\Pi $ has the following
property: for every $i$ if $t$ is a term in $p_{i}$, then $t$ is also in $p$.

Nienhuys-Cheng$^{[9]}$ used the above condition to discuss a metric on the
set of Herbrand interpretations for finite Horn logic programs. As we know,
in a constrained clause, it is usually assumed that every variable occurring
in the body of a clause also occurs in its head. In fact, assumption 4.2 can
be syntactically checked in polynomial time and easily satisfied. For example, suppose that there are one 
function symbol $f,$ one predicate symbol $p$ and one constant symbol $a$ in the base $L$.
It is easy to verify that there are infinitely many clauses satisfying this
assumption, e.g. $\pi _{k}=p(f^{k}(x))\leftarrow p(x)$, where $k=1,2,\cdots $%
. Based on this fact, we can conclude that there is an uncountable number of
infinte Horn logic programs satisfying assumption 4.2.

By the definition of the limits and the assumption that every $\Pi _{n}$ is
Horn, it is easily verified that $\Pi =\displaystyle\lim_{n\rightarrow
\infty }\Pi _{n}$ is Horn. Under assumption 4.2, we shall prove the
following theorem.

{\bf Theorem 4.3.} Let $L$ be a base consisting of infinitely many constant
symbols, function symbols and finitely many predicate symbols. Given a
sequence $\{\Pi _{n}\}$ of Horn logic programs in $L$, if $\Pi =\displaystyle%
\lim_{n\rightarrow \infty }\Pi _{n}$ exists and there is an $N$ such that $%
\Pi _{n}$ satisfies assumption 4.2 for every $n\geq N$, then $\displaystyle%
\lim_{n\rightarrow \infty }f_{\Pi _{n}}^{\omega }$ exists and $f_{\Pi
}^{\omega }=\displaystyle\lim_{n\rightarrow \infty }f_{\Pi _{n}}^{\omega }.$

{\sl Proof.} We first prove that $\overline{\displaystyle\lim }%
_{n\rightarrow \infty }f_{\Pi _{n}}^{\omega }\subseteq f_{\Pi }^{\omega }$.
Assume that $p\in \overline{\displaystyle\lim }_{n\rightarrow \infty }f_{\Pi
_{n}}^{\omega },$ then there is an infinite sequence $n_{i},i=1,2,\cdots ,$
such that for every $n_{i},p\in f_{\Pi _{n_{i}}}^{\omega }.$ Then by the
definition, for every $n_{i},$ there is a proof of $p$ from the facts of $%
\Pi _{n_{i}}.$ We define a proof tree $T$ of $p$ as follows such that every
node $\alpha $ of $T$ is associated with a set of atoms, denoted by $%
P(\alpha ).$ Let $\lambda $ be the root of $T,$ assume that $P(\lambda
)=\{p\}.$ For every clause $\pi \in \bigcup_{n=N}^{\infty }\Pi _{n}$ with
head $p,$ there is a node $\alpha \in T$ such that $\alpha $ is a child of $%
\lambda $ and $P(\alpha )=body(\pi ).$ For every node $\alpha \in T,$ if
there is a clause $\pi ^{\prime }\in \bigcup_{n=N}^{\infty }\Pi _{n}$ such
that $head(\pi ^{\prime })\in P(\alpha ),$ there is a child $\beta $ of $%
\alpha $ such that $P(\beta )=body(\pi ^{\prime })$ (if $\pi $ is a fact
then $P(\beta )=\emptyset $).

We construct a subtree $T^{\prime }$ of $T$ as follows: let $\lambda \in
T^{\prime },$ and let $\alpha $ be a child of $\lambda $ such that $%
p\leftarrow P(\alpha )$ is used to deduce $p$, by the assumption, such an $%
\alpha $ does exist, then set $\alpha \in T^{\prime };$ and for any node $%
\alpha ^{\prime }\in T^{\prime }$ and $p^{\prime }\in P(\alpha ^{\prime }),$
let $\beta $ be the child of $\alpha ^{\prime }$ such that $p^{\prime
}\leftarrow P(\beta )$ is used to deduce $p^{\prime }$, then set $\beta \in
T^{\prime }.$

We now prove that $T$ has only finitely many such subtrees. To do this, we
only need to show that $T$ is finite. Every ground atom has a tree structure
and the depth of the tree can be considered as the {\it level} of the atom$%
^{[9]}$. By assumption 4.2, the level of an atom in $T$ is not smaller than
that of any atom in its children. Hence, the level of every atom in $T$ is
not greater than that of $p$, i.e. the levels of atoms in $T$ are finite.
Also by assumption 4.2, every function symbol or constant symbol appearing
in the nodes of $T$ also appears in $p$. Note that there are only finitely
many predicate symbols. Thus $T$ is finite and so we are done.

It follows from the above analysis that there is a subtree $T^{\prime }$ of $%
T$ such that $T^{\prime }$ belongs to infinitely many $\Pi _{n}$'s. By the
definition of the limits, we have%
\[
T^{\prime }\subseteq \displaystyle\lim_{n\rightarrow \infty }\Pi _{n}. 
\]%
There is an $N_{0}$ and a $K$ such that for every $n\geq N_{0},\ T^{\prime
}\in f_{\Pi _{n}}^{K}.$ So $p\in f_{\Pi }^{K},$ i.e., $p\in f_{\Pi }^{\omega
}$. Note that $p\in \overline{\displaystyle\lim }_{n\rightarrow \infty
}f_{\Pi _{n}}^{\omega }$. Thus $\overline{\displaystyle\lim }_{n\rightarrow
\infty }f_{\Pi _{n}}^{\omega }\sqsubseteq f_{\Pi }^{\omega }$. By lemma 4.1, 
$f_{\Pi }^{\omega }\subseteq \underline{\displaystyle\lim }_{n\rightarrow
\infty }f_{\Pi _{n}}^{\omega }$. Recall that $\underline{\displaystyle\lim }%
_{n\rightarrow \infty }f_{\Pi _{n}}^{\omega }\sqsubseteq \overline{%
\displaystyle\lim }_{n\rightarrow \infty }f_{\Pi _{n}}^{\omega }$. So, we get

\[
\underline{\displaystyle\lim }_{n\rightarrow \infty }f_{\Pi _{n}}^{\omega }=%
\overline{\displaystyle\lim }_{n\rightarrow \infty }f_{\Pi _{n}}^{\omega
}=f_{\Pi }^{\omega }. 
\]%
This completes the proof. \hfill$\Box $

\medskip

The following discussion will point out an interesting decidability result
on the least Herbrand models related to the above main theorem. As we know,
given a finite Horn logic program, the least Herbrand model of this program
is in genreral non-recursive$^{[1]}$. In what follows, we will prove that
for finite Horn logic programs satisfying assumption 4.2, their least
Herbrand models are recursive. More precisely, we have the following theorem.

{\bf Theorem 4.4.} Given a finite Horn logic program $\Pi $, if $\Pi $
satisfies assumption 4.2, then the least Herbrand model of $\Pi $ is
recursive.

{\sl Proof.} For any ground atom $p$, using the clauses in $\Pi $, we can
construct a tree $T$ with $p$ as the root node by the same method as in the
proof of theorem 4.3. As previously shown, if $\Pi $ satisfies assumption
4.2, then $T$ is finite. Note that $T$ contains all the possible proof trees
for $p$ from $\Pi $. So we can decide in finite time whether $p$ is provable
from $\Pi $. Equivalently, it is decidable whether $p$ is in the least
Herbrand model of $\Pi $. This completes the proof. \hfill $\Box $

\bigskip

\noindent {\bf 5. The stability of fixed points for Horn logic programs}

\smallskip

For a Horn logic program $\Pi $, the Herbrand models of this program are
also the fixed points of the mapping $f_{\Pi }$. In the study of dynamical
systems, the concept of stability is of great importance to characterize the
properties of fixed points. Intuitively, the stability of a fixed point of a
differential equation means that if the initial solution is sufficiently
close to the fixed point then the solution will remain close to the fixed
point thereafter$^{[3]}$. There are many ways to formalize this concept
mathematically which are similar but not the same. A classical concept for
the stability of fixed points defined by Lyapounov is as follows$^{[3]}$:

A fixed point $X$ of $F:{\Bbb R}^{m}\rightarrow {\Bbb R}^{m}$ is {\it stable}
if for all $\epsilon >0$ there exists $\delta (\epsilon )$ such that

\[
\left| F^{n}(x_{0})-X\right| <\epsilon \text{ \ \ \ for }n=1,2,\cdots , 
\]%
for all $x_{0}$ such that $\left| x_{0}-X\right| <\delta (\epsilon )$, where 
$x_{n}=F^{n}(x_{0})$ is defined iteratively by $x_{n+1}=F(x_{n})$.

To discuss the stability of fixed points for Horn logic programs, it is
essential to define a distance between Herbrand interpretations. Fitting$%
^{[4]}$ was the first to introduce metric methods to logic programming. [4]
defined a metric on the set of Herbrand interpretations by level mappings
and studied the fixed point semantics for some logic programs using the
Banach contraction theorem. Nienhuys-Cheng$^{[9]}$ used the depth of an
expression tree for the level mapping and discussed this mapping and the
induced metric.

{\bf Definition 5.1.} \ ([9]) Let $HB$ be the Herbrand base of a logical
language. A level mapping is a function $\parallel :HB\rightarrow {\Bbb N}$
(natural numbers). We use $\left| A\right| $ to denote $\parallel \left(
A\right) $, i.e. the image of $A$. It is called the {\it level} of $A$.

Given a level mapping $\parallel $ we can define a metric on the set of all
Herbrand interpretations as follows: $d(I,I)=0$ for every interpretation $I$%
. If $I\neq J$, then $d(I,J)=1/2^{n}$ if $I$ and $J$ differ on some ground
atom of level $n$, but agree on all ground atoms of lower level$^{[4]}$,
i.e. $A\in I\cap J$ for all $\left| A\right| \leq n-1$ $\wedge $ $A\in I\cup
J$ and there is an $A\in I\vartriangle J$ such that $\left| A\right| =n$.

{\bf Theorem 5.2.} \ ([4]) The function $d$ defined above is a metric on the
set of all Herbrand interpretations $2^{HB}$.

{\bf Lemma 5.3.} \ ([4]) The metric space $\left( 2^{HB},d\right) $ is
complete.

{\bf Definition 5.4.} \ ([9]) Every ground atom has a tree structure and the
depth can be considered as the level of the atom. Thus we have a level
mapping $\parallel :HB\rightarrow {\Bbb N}$. Let $d$ be the distance on the
set of Herbrand interpretations induced by this level mapping.

For a Horn logic program $\Pi $, the mapping $f_{\Pi }$ is from the set of
Herbrand interpretations to the set of Herbrand interpretations. In this
paper, we will use the above metric to study the fixed points of Horn logic
programs. First, if a Horn logic program $\Pi $ satisfies assumption 4.2,
then the mapping $f_{\Pi }$ has the following property.

{\bf Lemma 5.5.} \ ([9]) Given a Horn logic program $\Pi $, if $\Pi $\
satisfies assumption 4.2, then $d(f_{\Pi }(I),f_{\Pi }(J))$ $\leq d(I,J)$
for all $I,J\in 2^{HB}$.

Given a Horn logic program $\Pi ,$ if there is a constant $c:0\leq c<1$ such
that $d(f_{\Pi }(I),f_{\Pi }(J))$ $<c\cdot d(I,J)$ for all $I,J\in 2^{HB}$,
then the mapping $f_{\Pi }$ is called a {\it contraction}. In such a case,
we know, by the Banach contraction theorem, that $f_{\Pi }$ has a unique
fixed point. For a Horn logic program $\Pi $ satisfying assumption 4.2, in
many cases the mapping $f_{\Pi }$ may be a contraction, e.g. $\Pi
=\{p(f(x))\leftarrow p(x),p(f(a))\}$, but not in general, e.g. $\Pi
=\{p(x)\leftarrow p(x),p(f(a))\}$ which has more than one fixed point such
as $\{p(f(a))\}$ and $\{p(a),p(f(a))\}$.

We can now define and discuss the stability of fixed points for Horn logic
programs.

{\bf Definition 5.6.} \ For a Horn logic program $\Pi $, a fixed point $J$
of $f_{\Pi }:2^{HB}\rightarrow 2^{HB}$ is {\it stable} if for all $\epsilon
>0$ there exists $\delta (\epsilon )$ such that

\[
d(f_{\Pi }^{n}(I),J)<\epsilon \text{ \ \ \ for }n=1,2,\cdots , 
\]%
for all interpretations $I$ such that $d(I,J)<\delta (\epsilon )$.

From the above definition, we can prove that the fixed points of Horn logic
programs satisfying assumption 4.2 are stable. That is to say, we have the
following theorem.

{\bf Theorem 5.7.} Given a Horn logic program $\Pi $, if $\Pi $ satisfies
assumption 4.2, and $J$ is a fixed point of $f_{\Pi }$, then $J$ is a stable
fixed point.

{\sl Proof.} To prove that $J$ is a stable fixed point, by definition 5.6,
we have to show that for any $\epsilon >0$ there exists $\delta (\epsilon )$
such that

\[
d(f_{\Pi }^{n}(I),J)<\epsilon \text{ \ \ \ for }n=1,2,\cdots , 
\]%
for all interpretations $I$ such that $d(I,J)<\delta (\epsilon )$.

Let $\delta (\epsilon )=\epsilon $. We can now prove the above inequality by
induction on $n$. When $n=0$, it is trivial to see that the inequality
holds. Assume that for $n=k$, the inequality also holds, i.e.

\[
d(f_{\Pi }^{k}(I),J)<\epsilon .\text{ \ \ } 
\]%
When $n=k+1$, applying lemma 5.2 yields

\[
d(f_{\Pi }^{k+1}(I),J)=d(f_{\Pi }^{k+1}(I),f_{\Pi }^{k+1}(J))\leq d(f_{\Pi
}^{k}(I),J)<\epsilon \text{,} 
\]%
as required. \hfill$\Box $

\smallskip

By theorem 5.7, it is easy to see that in example 3.3, the least Herbrand
model $M=\emptyset $ of the program $\Pi =\{p(f(x))\leftarrow p(x)\}$ is a
stable fixed point. But in example 3.2 where assumption 4.2 is not
satisfied, we can prove by contradiction that the least Herbrand model $%
M=\emptyset $ of the program $\Pi =\{p(x)\leftarrow p(f(x))\}$ is not a
stable fixed point. Suppose that $M=\emptyset $ is a stable fixed point. By
definition 5.6, for $\epsilon =1/8$, there exists a constant $\delta >0$
such that

\[
d(f_{\Pi }^{n}(I),M)<\frac{1}{8}\text{ \ \ \ for }n=1,2,\cdots , 
\]%
for all interpretations $I$ such that $d(I,M)<\delta $. Let $%
I_{0}=\{p(f^{k}(a))\}$ where $k\in {\Bbb N}$ and $2^{k}>1/\delta $. Then we
have

\[
d(I_{0},M)=\frac{1}{2^{k+2}}<\frac{1}{2^{k}}<\delta , 
\]%
and $f_{\Pi }^{k}(I_{0})=\{p(a)\}$. So $d(f_{\Pi }^{k}(I_{0}),M)=1/4$, which
contradicts the assumption that $d(f_{\Pi }^{n}(I_{0}),M)<1/8$ for $%
n=1,2,\cdots ,$ and thus completes the proof.

\smallskip

Theorems 4.3, 5.7 and examples 3.2, 3.3 show that there seems to exist some
connections between the limit of Horn logic programs and the stability of
fixed points for Horn logic programs. As mentioned above, a stable fixed
point is one where a small perturbation will not grow but remain close to
the fixed point even after an infinite number of iterations. So from this
point of view we can understand that if the least Herbrand model of a Horn
logic program is a stable fixed point, then this model may be approached by
a sequence of least Herbrand models.

\bigskip

\noindent {\bf 6. Conclusion}

\smallskip

Assume that the base $L$ contains infinitely many constant symbols, function
symbols and finitely many predicate symbols. Given a sequence $\{\Pi _{n}\}$
of Horn logic programs, suppose that the limit $\Pi $ of $\{\Pi _{n}\}$
exists, we studied if the least Herbrand model of $\Pi $ is equal to the
limit of the least Herbrand models of each $\Pi _{n}$. It was shown by some
examples that in general, this property is not true. Further studies proved
that the property holds if Horn logic programs satisfy an assumption that can be syntactically 
checked and easily satisfied. This
result implies that for an infinite Horn logic program satisfying the
assumption, the least Herbrand model of this program can be approached by a
sequence of least Herbrand models of finite programs. We also discussed the
decidability of least Herbrand models for Horn logic programs satisfying the
assumption. It is proved that the least Herbrand models of these programs
are recursive. More interestingly, using the concept of stability from
dynamical systems, we proved that the assumption is exactly a sufficient
condition to ensure the stability of fixed points for Horn logic programs.
Finally, we hope that more general conditions could be found for the property to hold 
and the main ideas and results of this paper could be
extended to more general logic programs in the future.

\bigskip

\noindent{\bf References:}

\begin{description}
\item \lbrack 1] Apt, K., Logic Programming, in J. Leeuwen (eds.), {\it %
Handbook of Theoretical Computer Science}, Elsevier Science, 1990, Chapter
10.

\item {[2]} Dahr, M., {\it Deductive Databases: Theory and Applications},
International Thomson Computer Press, 1997.

\item \lbrack 3] Drazin, P. G., {\it Nonlinear Systems, }Cambridge
University Press, 1992.

\item \lbrack 4] Fitting M., Metric methods, Three examples and a Theorem, 
{\it J. of Logic Programming}, 21(1994):113-127.

\item {[5]} Li, W., An Open Logic System, {\it Science in China} (Scientia
Sinica) (series A), 10(1992)(in Chinese), 1103-1113.

\item {[6]} Li, W., A Logical Framework for Evolution of Specifications, in: 
{\it Programming Languages and Systems,}\ (ESOP'94), LNCS 788,
Sringer-Verlag, 1994, 394-408.

\item {[7]} Li, W., A logical Framework for Inductive Inference and Its
rationality, in N. Fu (eds.), {\it Advanced Topics in Artificial
Intelligence,} LNAI 1747, Springer, 1999.

\item \lbrack 8] Nienhuys-Cheng, S. H., Distance between Herbrand Interpretations: 
A Measure for Approximations to a Target Concept, in: {\it Proc. of the 7th International
Workshop on Inductive Logic Programming}, LNAI 1297, Springer, 1997,
213-226. 

\item {[9]} Nienhuys-Cheng, S. H., Distances and Limits on Herbrand
Interpretations, in: {\it Proc. of the 8th International Workshop on Inductive
Programming}, LNAI 1446, Springer, 1998, 250-260.

\item {[10]} van Emden, M. H. and Kowalski, R. A., The Semantics of
Predicate Logic as a Programming Language, {\it J. Association for Computing
Machinary} 23(4)(1976), 733-742.
\end{description}

\end{document}